\documentclass[prd,twocolumn,floatfix,amsmath,nofootinbib,amssymb,floatfix]{revtex4}
\usepackage{graphicx,color,dcolumn,booktabs,bm,multirow}
\usepackage{longtable,lscape}
\usepackage{txfonts}
 \usepackage{enumitem}
\usepackage{overpic}
\usepackage{amssymb}
\usepackage{indentfirst}
\usepackage{feynmf}   
\usepackage{slashed}  
\usepackage{cases}
\usepackage{color}
\usepackage{multirow}
\usepackage{epstopdf}
\usepackage{graphicx,color,dcolumn,booktabs,bm}

\usepackage[colorlinks,
            citecolor=blue,
            anchorcolor=red,
            menucolor=red,
            linkcolor=red,
            filecolor=red,
            runcolor=red,
            urlcolor=blue,
            frenchlinks=red]{hyperref}

\begin{document}
\title{Radiative decays of the charmed mesons in a modified relativistic quark model}
\author{Jie-Lin Li$^1$}
\author{Dian-Yong Chen$^{1,2}$} \email{chendy@seu.edu.cn}
\affiliation{
$^1$ School of Physics, Southeast University,  Nanjing 210094, China\\
$^2$ Lanzhou Center for Theoretical Physics, Lanzhou University, Lanzhou 730000, China\\
}
\date{\today}
\begin{abstract}
In the present work, we perform systematic estimations for the radiative decays of the charmed mesons in a modified relativistic quark model. Our estimations indicate that the branching ratios of the process of  $D_2^0(1^3P_2) \to D^{\ast 0}(1^3S_1) \gamma$,  $D_3^0(1D_3) \to D_2^0(1^3P_2) \gamma$, $D_2^0(2D_2^\prime) \to D_1^{0}(2P_1) \gamma$, $D_3^0(2^3D_3) \to D_2^0(2^3P_2) \gamma$, and $D^{\ast 0}(1^3S_1) \to D^0(1^1S_0) \gamma$ are of the order of $10^{-2}$, which are sizable to be detected experimentally. Moreover, the branching ratios of some channels, for example, $D_1^0(1P_1) \to D(1^1S_0)^0 \gamma$, $D^0(3^1S_0) \to D_1^{\prime 0}(2P^\prime_{1}) \gamma$ and $D^0(3^3S_1) \to D_2^0(2^3P_2) \gamma$ are estimated to be of the order of $10^{-3}$, which may be accessible with the accumulation of data in future experiments.
\end{abstract}
\maketitle
\section{Introduction}

After the observations of $J/\psi$ in 1974 \cite{SLAC-SP-017:1974ind,E598:1974sol}, the crucial test of the notion of the charm quark is the existence of charmed mesons near 2 GeV as indicated in Ref.~\cite{DeRujula:1974rkb}. Many experimental attempts had been performed in various processes and an increasing number of charmed mesons have been observed. In the following, we present a short review of the discovery history of the charmed mesons.

\begin{itemize}[leftmargin=*]
\item {\it The lowest $S$ wave charmed mesons.} The first charmed meson, $D^{0}$, was observed after two years of the observation of $J/\psi$ in the $K^\pm \pi^\mp$ and $K^\pm \pi^\mp \pi^\pm \pi^\mp$ invariant mass spectra produced by $e^{+}e^{-}$ annihilation at the center-of-mass energies between 3.90 GeV and 4.60 GeV~\cite{Goldhaber:1976xn}, and soon later its charged partner $D^{\pm}$ was discovered in the similar processes~\cite{Peruzzi:1976sv}. In the year of 1977, another charmed meson, $D^\ast$ was observed in the $e^{+}e^{-}$ annihilation process~\cite{Goldhaber:1977qn}, which has been identified as $1^{3}S_{1}$ charmed meson. Then the lowest $S$ wave charmed mesons were well established.
\item {\it The lowest $P$ wave charmed mesons.} As for $P$ wave charmed mesons, the $J^P=1^+$ state in the spin triplet can mix with the one in the spin singlet. In the heavy quark limit, one physical $1^+$ state dominantly couples to $D^\ast \pi$ via $S$ wave, while another one dominantly couples to $D^\ast \pi$ via $D$ wave. In this case, the former $D_1$ should be much broader than the latter one. On the experimental side, the first $P$ wave charmed meson, $D_{1}(2420)$, was observed in 1986 by the ARGUS collaboration at the DORIS II $e^{+}e^{-}$ storage ring at DESY~\cite{ARGUS:1989mcc}. The mass and width were measured to be $2420\pm 6$ MeV and $70 \pm 21$ MeV, respectively~\cite{ARGUS:1989mcc}. The angular momentum analysis indicated that the $J^P$ quantum numbers of this state should be $1^{+}$~\cite{ARGUS:1989mcc}, which implies that $D_1(2420)$ should be a $P$ wave charmed meson. After the discovery of $D_{1}(2420)$, it had been further confirmed by some other measurements~\cite{TaggedPhotonSpectrometer:1988qan, DELPHI:1998oyl, CDF:2005zry, ZEUS:2008nzg, BaBar:2010zpy, LHCb:2013jjb, LHCb:2019juy}. The second observed $P$ wave charmed meson is $D_2(2460)$, which was observed in the $D^{+}\pi^{-}$ invariant mass spectrum by E691 experiment in Fermilab~\cite{Anjos:1988uf}. The resonance parameters were observed to be $m=2459 \pm 3$ MeV and $\Gamma=20 \pm 10 \pm 5$ MeV, respectively, and its spin-parity is determined as $J^{P}=2^{+}$.

 The broader $D_1$ state, $D_{1}(2430)$, with a mass $2427\pm 26 \pm 20 \pm 15$ MeV and a width $384^{+107}_{-75} \pm 24 \pm 70$ MeV was detected by Belle Collaboration in 2004~\cite{Abe:2003zm}. Besides $D_1(2430)$, another broad state $D_{0}^{*}(2400)$ was observed in the $D \pi$ invariant mass spectrum of $B \to D \pi \pi$ process, which can be categorized as the last lowest $P$ wave charmed meson with $J^P=0^+$. The measure mass and width are $2308 \pm 17 \pm 15 \pm 28$ and $276 \pm 21\pm 63$ MeV, respectively. Then all the lowest $P$ wave charmed meson had been observed.

\item {\it $D_0(2550)$ and $D_1^\ast (2600)$ as $2S$ charmed mesons candidates.} In 2010, the BABAR Collaboration performed an analysis of the $D^+ \pi^-$, $D^0 \pi^+$, and $D^\ast \pi^- $ system in the inclusive $e^ +e^- \to c\bar{c} $ interaction to search for new excited charmed mesons, and four new states were observed, which are  $D_0(2550)$, $D_1^\ast (2600)$, $D_3(2750)$, and $D_1^\ast(2760)$~\cite{delAmoSanchez:2010vq}. The mass and width of $D_0(2550)$ were measured to be $2539.4\pm4.5\pm6.8$ and $130 \pm 12 \pm 13$ MeV, respectively, and the $J^P$ quantum numbers were determined to be $0^-$, which indicates $D_0(2550)$ could be a good candidate of $D(2^1S_0)$ state. The estimations of the mass spectrum and decay properties in Refs. \cite{Godfrey:1985xj, Chen:2011rr, Lu:2014zua} supported the $D(2^1S_0)$ assignment for $D_0(2550)$.

As for $D_1^\ast (2600)$, the resonance parameters were measured to be $m=2608.7 \pm 2.4 \pm 2.5$ MeV and $\Gamma = 93 \pm 6 \pm 13$ MeV, respectively, while the helicity angle distributions indicated that its $J^P$ quantum numbers were $1^-$. The estimation in Ref.~\cite{Li:2010vx} indicated that the mass of $D_1^{*}(2600)$ was very close to the predicted one for the $D(2^{3}S_{1})$. The investigations of strong decay behaviors in the heavy quark effective theory with the leading order approximations supported that the  $D_1^\ast(2600)$ could be the first  radial excitation of the $D^{*}$. It is also worth to mention that the estimations in Refs.~ \cite{Yu:2014dda, Zhong:2010vq, Sun:2010pg, Lu:2014zua} also identified the $D_1^{*}(2600)$ as  $2^{3}S_{1}$ charmed meson.

\item {\it $D_3(2750)$, $D_1^\ast (2760)$, and $D_2(2740)$ as $D$ wave charmed meson candidates.}
As indicated in Ref.~\cite{delAmoSanchez:2010vq}, $D_1^\ast(2760)$ and $D_3(2750)$ were observed in the invariant mass spectra of $D \pi$ and $D^\ast \pi$, respectively, and their mass and widths differ by $2.6~\sigma$ and $1.5~\sigma$, respectively. The Dalitz plot analysis of $B^{0}\rightarrow \bar{D}^{0}\pi^{+}\pi^{-}$ performed by LHCb collaboration~\cite{Aaij:2015sqa} indicated that spin-parity of $D_3(2750)$ should be $3^-$, which was a good candidate of $1^3D_3$ charmed meson. As for $D_1^\ast (2760)$, the analysis by LHCb collaboration indicated that its spin-parity should be $1^-$. The estimations in the frame of relativistic quark model in Ref. \cite{DiPierro:2001dwf} indicated that $D_1^\ast (2760)$ could be a good candidate of $1^{3}D_{1}$ charmed meson. By investigating the mass spectrum and decay properties of charmed mesons, the authors in Ref.~\cite{Sun:2010pg} explained $D_1^\ast (2760)$ as a mixed state of $D(2^{3}S_{1})$ and $D(1^{3}D_{1})$.

Similar to the case of $P$ wave charmed mesons, the two physical charmed mesons with $J^P=2^-$ are the mixtures of the spin singlet and triplet. The first $2^-$ charmed meson, $D_{2}(2740)$, was observed in the $D^{*}\pi$ invariant mass spectrum of  the inclusive reaction $pp \to D\pi X$ by the LHCb collaboration in 2013~\cite{LHCb:2013jjb}. The mass and width were observed to be $2737.0\pm3.5\pm11.2$ MeV  and $73.2\pm13.4\pm25.0$ MeV, respectively. Further analysis of the helicity angle distributions of $B^{-}\rightarrow D^{*+}\pi^{-}\pi^{-}$ indicates the spin-parity if $D_2(2740)$ are $2^-$~\cite{LHCb:2019juy}, which imply $D_{2}(2740)$ should be a $1^3D_2$ charmed meson.

\item {\it Higher charmed meson around 3 GeV.} In 2013 the LHCb Collaboration observed two new charmed meson, $D_J(3000)$ and $D^{*}_{J}(3000)$, in the $D^{*}\pi$ and $D\pi$ invariant mass spectrum of the inclusive reactions $pp\to D^{(\ast)} \pi X$, respectively \cite{Aaij:2013sza}. The resonance parameters of $D_J(3000)$ are $m=2971.8\pm8.7$ MeV and $\Gamma=188.1\pm44.8$ MeV, respectively. The helicity angular distributions of $D_J(3000)$ is compatible with unnatural parity. As for $D^{*}_{J}(3000)$, its mass and width are $3008.1\pm4.0$ MeV  and $110.5\pm11.5$ MeV, respectively \cite{Aaij:2013sza}. By further analysis of the $D \pi $ invariant mass spectra of the exclusive process $B^{-}\rightarrow D^{+}\pi^{-}\pi^{-}$, the LHCb collaboration find one more broad resonance in the $D\pi$ invariant mass distribution above 3 GeV~\cite{Aaij:2016fma}. The resonance parameters were fitted to be $M=3214\pm29\pm49$ MeV, $\Gamma=186\pm38\pm34$ MeV, respectively, and the Dalitz plot analysis indicated that its spin-parity were $2^+$~\cite{Aaij:2016fma}. The measured resonance parameters of $D_2^\ast (3000)$ were not consistent with those of $D_J^\ast(3000)$, which indicates the origins of $D_2^\ast(3000)$ and $D_J^\ast(3000)$ should be different.

Besides the $1F$ charmed mesons, the masses of $2P$ and $3S$ charmed mesons were also predicted to be around 3 GeV ~\cite{Godfrey:1985xj}, thus, these states can be good candidates of $1F$ and $2P$ charmed mesons. In Ref. ~\cite{Yu:2014dda}, the authors investigated the decay properties of $D_J(3000)$ and $D^{*}_{J}(3000)$ by using $3P_{0}$ model and they concluded that $D_J(3000)$ could be categorized as first radial excitation of $D_{1}(2430)$, while $D^{*}_{J}(3000)$ as $D(1^{3}F_{2})$ or $1^{3}F_{4}$ charmed meson. The estimation in Ref. \cite{Lu:2014zua} indicated that $D^{*}_{J}(3000)$ could be assigned as $1^{3}F_{4}$, but $D_J(3000)$ was a good candidate of $D(3^{1}S_{0})$ charmed meson .
\end{itemize}

\begin{figure}[htb]
\centering
\includegraphics[width=80mm]{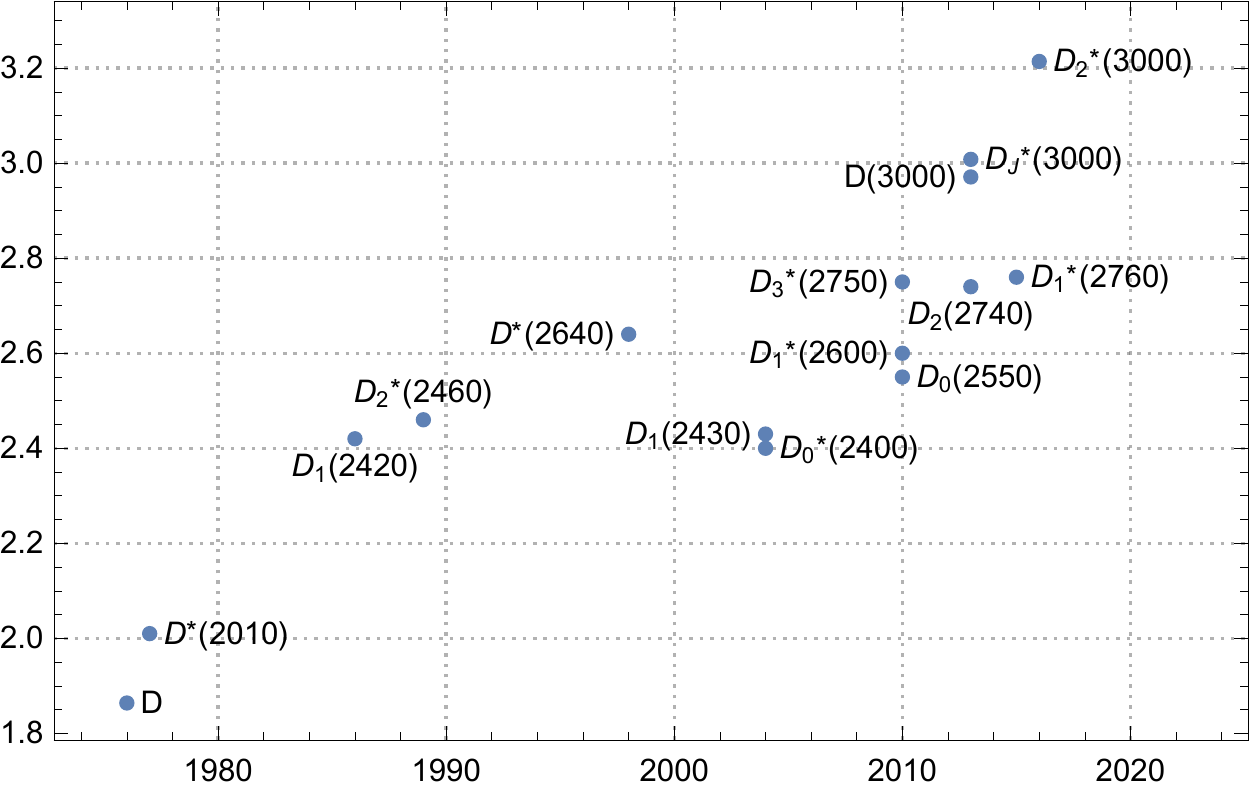}
\caption{ The history of charm meson discoveries~\cite{Peruzzi:1976sv,Albrecht:1985as,Anjos:1988uf,Abe:2003zm,delAmoSanchez:2010vq,Aaij:2013sza,Aaij:2015vea,Aaij:2016fma}. Here, the masses are taken from the Particle Data Group.\label{Fig:His}}
\end{figure}

A sketch diagram of the charmed mesons discovery history is presented in Fig.~\ref{Fig:His}, where one can find most higher charmed mesons were observed in the past decade. With the high energy and high luminosity beams at the LHC and SuperKEKB, higher charmed mesons have been observed, which undoubtedly enrich the charmed meson spectrum. The investigations of their properties are crucial for us to properly categorize these newly observed charmed mesons. Besides the strong decay process, the electromagnetic decays of the hadron are sensitive to its inner structure, so the radiative transitions can also probe the internal charge structure of mesons and be useful in determining meson structure.

To investigate the radiative decay properties of charmed mesons, we employ the quark model to estimate the mass spectra and wave functions. In Ref.~\cite{Godfrey:1985xj}, Godfrey and Isgur proposed a relativistic quark model (GI model) with chromodynamics to describe the whole meson spectra. The GI model achieved great success in describing the ground state of the meson spectra. However, the model failed to deal with the higher excited states due to the simple linear dconfinement potential without including the unquenched effects, which are expected to be important for higher excited states~\cite{Li:2009ad}. In Ref.~ \cite{Song:2015nia}, we replace the simple linear potential with the screened potential, which can in a certain extend reflect the unquenched effects in the higher excited charmed mesons. With the modified GI model~\cite{Song:2015nia}, we find the mass spectra and the strong decay properties of the higher charmed mesons could be better described than the GI model. In the present work, we employ the modified GI model to investigate the radiative decay properties of the charmed mesons.

This work is organized as follows. After the introduction, we present a review of the quark model employed in the present estimation and the formulas for radiative decays. The numerical results and discussions are given in section III, and the last section is devoted to a short summary.

\section{Radiative decay of charmed mesons}

\subsection{GI quark model}
From the perspective of classical physics, the light quark in the heavy-light meson system moves very fast relative to the heavy quark. Therefore, the relativistic effect cannot be ignored when dealing with heavy-light meson systems. Godfrey and Isgur proposed the relativistic quark model in 1985~\cite{Godfrey:1985xj}, which can well describe the ground state meson spectrum of heavy-light meson system. The hamiltonian can be expressed as,
\begin{eqnarray} \label{one}
\tilde{H}&=&\left(p^{2}+m^{2}_{1}\right)^{1/2}+\left(p^{2}+m^{2}_{2}\right)^{1/2}+\tilde{V}_{\mathrm{eff}}\left(\mathbf{p},\mathbf{r}\right),
\end{eqnarray}
where $\tilde{V}_{\mathrm{eff}}(\mathbf{p},\mathbf{r})=\tilde{V}^{\mathrm{conf}}+\tilde{V}^{\mathrm{hyp}}+\tilde{V}^{\mathrm{SO}}$ is the effective confinement potential between the quark and anti-quark in a meson. In the non-relativistic approximation, it can be expressed as
\begin{eqnarray}
V_{\mathrm{eff}}(r)&=&V^{\mathrm{conf}}(r)+V^{\mathrm{hyp}}(r)+V^{\mathrm{SO}}(r),
\end{eqnarray}
the spin-independent term $\tilde{V}^{\mathrm{conf}}$ is mainly composed of two parts. A short-distance $\gamma^{\mu}\otimes\gamma_{\mu}$ interaction of one-gluon-exchange and a long-distance $1\otimes1$ linear confining interaction. The specific forms of these two interactions are a Coulomb potential and a long-range linear potential,
\begin{eqnarray}
V^{\mathrm{conf}}_{ij}(r)&=&-\left[\frac{3}{4}c+\frac{3}{4}br-\frac{\alpha_{s}(r)}{r}\right]\left(\mathbf{F}_{i}\cdot\mathbf{F}_{j}\right), \label{Eq:Vconf}
\end{eqnarray}
$i$ and $j$ represent different quarks in the system respectively. The concrete form of the color-hyperfine term is
\begin{eqnarray}
V^{\mathrm{hyp}}(r)&=&-\frac{\alpha_{s}(r)}{m_{1}m_{2}}\left[\frac{8\pi}{3}\mathbf{S}_{1}\cdot\mathbf{S}_{2}\delta^{3}(\mathbf{r})\right.
\nonumber\\
& &\left. {}+\frac{1}{r^{3}}\left(\frac{3\mathbf{S}_{1}\cdot\mathbf{rS}_{2}\cdot r}{r^{2}}-\mathbf{S}_{1}\cdot\mathbf{S}_{2}\right)\right]\Big(\mathbf{F}_{i}\cdot\mathbf{F}_{j}\Big).
\end{eqnarray}
The spin orbit term $V^{\mathrm{SO}}$ consists of two parts and it reads
\begin{eqnarray}
V^{\mathrm{SO}}_{ij}(r)&=&V^{\mathrm{SO(cm)}}_{ij}+V^{\mathrm{SO(tp)}}_{ij},
\end{eqnarray}
where the former one is the color-magnetic term and the later one is Thomas-precession term. The concrete forms are as follows,
\begin{eqnarray}
V^{\mathrm{SO(cm)}}_{ij}(r)&=&-\frac{\alpha_{s}(r)}{r^{3}}\left[\frac{1}{m_{i}}+\frac{1}{m_{j}}\right]\left[\frac{\mathbf{S}_{i}}{m_{i}}+\frac{\mathbf{S}_{j}}{m_{j}}\right]\cdot\mathbf{L}\Big(\mathbf{F}_{i}\cdot\mathbf{F}_{j}\Big),\nonumber\\ \nonumber\\
V^{\mathrm{SO(tp)}}_{ij} (r)& =&-\frac{1}{2r}\frac{\partial V^{\mathrm{conf}}_{ij}}{\partial r}\left[\frac{\mathbf{S}_{i}}{m^{2}_{i}}+\frac{\mathbf{S}_{j}}{m^{2}_{j}}\right]\cdot\mathbf{L}
\label{Eq:Vsotp},
\end{eqnarray}
where $\mathbf{S}_{1}$ and $\mathbf{S}_{2}$ represent the spins of the quark and anti-quark, respectively. While $\mathbf{L}=\mathbf{r}\times\mathbf{p}$ stands for the orbital angular momentum. As for $\mathbf{F}_i \cdot \mathbf{F}_j$, its value is $-4/3$ in meson systems.

The influence of the relativistic effect is considered from two aspects. Firstly, the non-locality of the effective potential of the quark and antiquark interactions and the momentum dependences of the effective potential on momentum. Considering the above two kinds of relativistic effects. For the first kind of correction, on can firstly introduce a smearing function, which is
\begin{eqnarray}
\rho_{12}(\mathbf{r}-\mathbf{r'})&=&\frac{\sigma^{3}_{12}}{\pi^{3/2}}\mathrm{exp}\left(-\sigma^{2}_{12}\left(\mathbf{r}-\mathbf{r'}\right)\right)
\end{eqnarray}
with
\begin{eqnarray}
\sigma^{2}_{12}&=&\sigma^{2}_{0} \left[\frac{1}{2}+\frac{1}{2}\left(\frac{4m_{1}m_{2}}{(m_{1}+m_{2})^{2}}\right)^{4}\right]+s^{2}\left(\frac{2m_{1}m_{2}}{m_{1}+m_{2}}\right)^{2}
\end{eqnarray}
The potentials relativistic form is
\begin{eqnarray}
\tilde{f_{12}}(r)&=&\int d^{3}\mathbf{r'}\rho_{12}\left(\mathbf{r}-\mathbf{r'}\right)f\left(r'\right),
\end{eqnarray}
where $\tilde{f}(r)$ and $f(r))$ represent the potentials in the relativistic quark model and those in the nonrelativistic quark model, respectively.

The second kind of relativistic corrections is introducing the momentum dependences in the effective potentials. The corrected Coulomb potential is expressed as~\cite{Godfrey:1985xj}:
\begin{eqnarray}
\tilde{G}(r)\rightarrow\left[1+\frac{p^{2}}{E_{1}E_{2}}\right]^{1/2}\tilde{G}(r)\left[1+\frac{p^{2}}{E_{1}E_{2}}\right]^{1/2}
\end{eqnarray}
The corrected forms of tensor potential, contact potential, vector spin-orbit potential and scalar spin-orbit potential are as follows:
\begin{eqnarray}
\frac{\tilde{V}_{i}(r)}{m_{1}m_{2}}\rightarrow \left[1+\frac{p^{2}}{E_{1}E_{2}}\right]^{1/2+\epsilon_{i}}\frac{\tilde{V}_{i}(r)}{m_{1}m_{2}}\left[1+\frac{p^{2}}{E_{1}E_{2}}\right]^{1/2+\epsilon_{i}}
\end{eqnarray}
For different effective potentials, the parameters $\epsilon_i$ are different, which have been listed in Ref.~\cite{Godfrey:1985xj}.

\subsection{Modified GI quark model}
GI model is successful in describing the low-lying meson spectrum, but it has encountered problems in describing highly excited states. The predictions given by the GI model is quite different from the experimental measurement~\cite{BaBar:2003oey,CLEO:2003ggt,Belle:2003kup,BaBar:2006eep,Born:1989iv}, for example, the observed masses of $X(3872)$, $D_{s0}^\ast (2317)$ and $D_{s1}(2460)$ are far below the expectations the relativistic quark model~\cite{ParticleDataGroup:2012pjm,Godfrey:1985xj}. Further theoretical estimations indicated that the near-threshold effect of the coupling channel will depress the mass of the high excited-state~\cite{vanBeveren:2003kd,Dai:2006uz,Liu:2009uz}. From the perspective of quark model, the coupling channel effects can be phenomenologically described by screening the color charges at large distance, for example, the distance greater than 1 fm, by creating light quark-antiquark pairs. The estimations by using the unquenched Lattice QCD~\cite{Bali:2005fu,Armoni:2008jy,PACS-CS:2011ngu} and holographic models~\cite{Bigazzi:2008gd} have confirmed the existence of the screening effects in hadrons.

In the literatures, the screening effect in  charmonia~\cite{Li:2009ad,Li:2009zu}, light unflavored mesons~\cite{Mezoir:2008vx} and heavy light mesons~\cite{Song:2015nia, Song:2015fha} have been investigated by flattening the linear confinement potential. In Ref.~\cite{Li:2009ad}, the screening effect in charmonia was proposed to be described by replacing the linear confinement potential with a screened potential in the form,
\begin{eqnarray}
br\rightarrow V^{\mathrm{scr}}(r)=\frac{b(1-e^{-\mu r})}{\mu},
\end{eqnarray}
where the corrected potential behaves as a linear potential $br$ at the short distance and approaches a constant $b/\mu$ at a long distance. Following the method in Ref.~~\cite{Godfrey:1985xj}, one can transfer the non-relativistic effective potential to the relativistic form by,
\begin{eqnarray}
\tilde{V}^{\mathrm{scr}}&=&\int d^{3}\mathbf{r'}\rho_{12}(\mathbf{r}-\mathbf{r'})\frac{b(1-e^{-\mu r'})}{\mu}.
\end{eqnarray}

It should be noticed that the above screened potential scheme was also introduced to describe the heavy light meson families in Refs.~\cite{Song:2015nia,Song:2015fha}, where the mass spectrum and strong decay behaviors of excited charmed and charmed strange mesons can be well described.
In the present work, we adopt the same modified GI model as the one in Ref.~\cite{Song:2015fha} to estimate the mass spectrum and radiative decays of the charmed mesons. The estimated mass spectrum of charmed mesons are listed in Table ~\ref{Tab:Mass}. For comparison, we also listed the spectrum predicted by the GI model and the experimental measurements~\cite{ParticleDataGroup:2012pjm,delAmoSanchez:2010vq}. From the table, one can find the modified GI model can better describe the mass spectrum of charmed mesons.

\begin{table}[htbp]
\caption{Spectrum of the charm mesons in unit of MeV. The theoretical predictions of GI model and the experimental measurements are also presented for comparison. \label{Tab:Mass}}
\centering
\begin{tabular}{cccc}
\toprule[1pt]
States & GI model \cite{Godfrey:1985xj} & Modified GI model \cite{Song:2015fha} & PDG \cite{ParticleDataGroup:2012pjm,delAmoSanchez:2010vq} \\
\midrule[0.5pt]
$1^{1}S_{0}$ & 1874 & 1861 & $1864.84\pm0.07$ \\
$2^{1}S_{0}$ & 2583 & 2534 & $2539.4\pm4.5\pm6.8$ \\
$3^{1}S_{0}$ & 3068 & 2976 & \\
$1^{3}S_{1}$ & 2038 & 2020 & $2010.26\pm0.7$ \\
$2^{3}S_{1}$ & 2645 & 2593 & $2608.7\pm2.4\pm2.5$ \\
$3^{3}S_{1}$ & 3111 & 3015 & \\
\midrule[0.5pt]
$1P_{1}$ & 2455 & 2424 & $2421.4\pm0.6$ \\
$2P_{1}$ & 2933 & 2866 & \\
$1^{3}P_{0}$ & 2398 & 2365 & $2318\pm29$ \\
$2^{3}P_{0}$ & 2932 & 2865 & \\
$1P_{1}^{ \prime }$ & 2467 & 2434 & $2427\pm26\pm25$ \\
$2P_{1}^{ \prime }$ & 2952 & 2872 & \\
$1^{3}P_{2}$ & 2501 & 2468 & $2464.3\pm1.6$ \\
$2^{3}P_{2}$ & 2957 & 2884 & \\
\midrule[0.5pt]
$1D_{2}$ & 2827 & 2775 & \\
$2D_{2}$ & 3225 & 3127 & \\
$1^{3}D_{1}$ & 2816 & 2762 & \\
$2^{3}D_{1}$ & 3231 & 3131 & \\
$1D_{2}^{ \prime }$ & 2834 & 2777 & \\
$2D_{2}^{ \prime }$ & 3235 & 3133 & \\
$1^{3}D_{3}$ & 2833 & 2779 & \\
$2^{3}D_{3}$ & 3226 & 3129 & \\
\bottomrule[1pt]
\end{tabular}
\end{table}

\subsection{Formalism for radiative decays}
With the wave function evaluated by the modified GI model, one can investigate the radiative decays of the charmed mesons. The start point of the radiative decay is the quark-photon electromagnetic coupling at tree level, which is,
\begin{eqnarray}
H_{em}&=&-\sum_{j}e_{j}\bar{\psi}_{j}\gamma^{j}_{\mu}A^{\mu}(\mathbf{k},\mathbf{r})\psi_{j}
\end{eqnarray}
where ${\psi}_{j}$ represents the $j$-th quark field in the  charm meson and $e_{j}$ is the charge carried by the $j$-th constituent quark. The $\mathbf{k}$ is the three momentum of the emitted photon.

In the non-relativistic limit, the spinor $\bar{\psi}$ can be replaced by $\psi^{\dag}$, while the $\bm{\gamma}$ matrices turn into   $\bm{\alpha}$ matrix. In this case, the electromagnetic transitions operator can be expressed as~\cite{Li:1997gd, Close:1970kt, Li:1994cy},
\begin{eqnarray}
\label{two}
h_{e}\simeq\sum_{j}[e_{j}\mathbf{r}_{j}\cdot\bm{\epsilon}-\frac{e_{j}}{2m_{j}}\bm{\sigma}_{j}\cdot(\bm{\epsilon}\times\mathbf{\hat{k}})]e^{-i\mathbf{k}\cdot \mathbf{r}_{j}}
\end{eqnarray}
where $m_{j}$ and $\mathbf{r}_{j}$ are the mass and coordinate of the $j$-th constituent quark, respectively, while $\bm{\sigma}_{j}$ is the Pauli matrix and $\mathbf{\epsilon}$ is polarization vector of the photon. In the above transition operator, the first and the second terms correspond to the electric and magnetic transitions, respectively.

Considering the electromagnetic transitions between the initial state $|i\rangle$ and the final state $|f\rangle$, the helicity amplitudes of the electric and magnetic transitions read,
\begin{eqnarray}
\mathcal{A}^{E}_{\lambda}&=&-i\sqrt{\frac{\omega_{\gamma}}{2}}\bigg\langle f \Big|\sum_{j}e_{j}\mathbf{r}_{j}\cdot\bm{\epsilon}e^{-i\mathbf{k}\cdot\mathbf{r}_{j}} \Big| i \bigg\rangle, \\
\mathcal{A}^{M}_{\lambda}&=&+i\sqrt{\frac{\omega_{\gamma}}{2}}\bigg\langle f \Big|\sum_{j}\frac{e_{j}}{2m_{j}}\bm{\sigma}_{j}\cdot(\bm{\epsilon}\times\bm{\hat{k}})e^{-i\mathbf{k}\cdot\mathbf{r}_{j}} \Big| i \bigg \rangle,
\label{Eq:AmpHel1}
\end{eqnarray}
where $\omega_{\gamma}$ is energy of the emitted photon.

In the initial hadron stationary system, one can choose the photon momentum $\mathbf{k}$ along the $z$ axial direction, i.e., $\mathbf{k}=\{0,0,k\}$. and polarization vector of the photon with the right-hand to be $\bm{\epsilon}=-\frac{1}{\sqrt{2}}(1,i,0)$. Considering the multipole expansion of $e^{-i\mathbf{k}\cdot\mathbf{r}_{j}}$, the helicity amplitudes in Eq. (\ref{Eq:AmpHel1}) can be expanded by amplitudes with a certain angular momentum $l$, which are~\cite{Deng:2016stx, Chen:2020jku},
\begin{eqnarray}
\mathcal{A}^{El}_{\lambda}&=&\sqrt{\frac{\omega_{\gamma}}{2}}\bigg\langle J^{\prime}\lambda^{\prime} \Big| \sum_{j}(-i)^{l}B_{l}e_{j}j_{l+1}(kr_{j})r_{j}Y_{l1}\Big| J\lambda \bigg \rangle
\nonumber\\
& & {}+\sqrt{\frac{\omega_{\gamma}}{2}} \bigg\langle J^{\prime}\lambda^{\prime} \Big| \sum_{j}(-i)^{l}B_{l}e_{j}j_{l-1}(kr_{j})r_{j}Y_{l1}\Big| J\lambda \bigg\rangle , \nonumber
\end{eqnarray}
\begin{eqnarray}
\mathcal{A}^{Ml}_{\lambda}&=&\sqrt{\frac{\omega_{\gamma}}{2}}\bigg\langle J^{\prime}\lambda^{\prime}\Big|\sum_{j}(-i)^{l}C_{l}\frac{e_{j}}{2m_{j}}j_{l-1}(kr_{j})
\nonumber\\
& & {}\times[\sigma^{+}_{j}\otimes Y_{l-10}]^{l}_{1}\Big|J\lambda\bigg\rangle
\nonumber\\
& & {}+\sqrt{\frac{\omega_{\gamma}}{2}}\bigg\langle J^{\prime}\lambda^{\prime}\Big|\sum_{j}(-i)^{l}C_{l}\frac{e_{j}}{2m_{j}}j_{l-1}(kr_{j})
\nonumber\\
& & {}\times[\sigma^{+}_{j}\otimes Y_{l-10}]^{l-1}_{1}\Big|J\lambda\bigg\rangle,
\end{eqnarray}
with $B_{l}=\sqrt{2\pi l(l+1)/(2l+1)}$ and $C_{l}=i\sqrt{8\pi (2l-1)}$.

As indicated in Ref.~\cite{Chen:2020jku}, the high angular momentum contributions are neglectable comparing to the lowest order approximation with $l=1$. Thus, in the lowest order approximation, the electric transition widths is~\cite{Godfrey:2016aq},
\begin{eqnarray}
\label{Eq:E1}
\Gamma^{E_1}&=&\frac{4}{3}\alpha \bigg(\frac{e_{1}m_{2}}{m_{1}+m_{2}}-\frac{e_{2}m_{1}}{m_{1}+m_{2}}\bigg)^{2}\omega^{3}_{\gamma}\delta_{SS^{\prime}}\delta_{LL^{\prime}\pm1}\max\left(L,L^{\prime}\right)
\nonumber\\
& & {}(2J^{\prime}+1)\begin{Bmatrix}L^{\prime} & J^{\prime} & S \\ J & L & 1\end{Bmatrix}^{2}\left\langle n^{\prime 2S^{\prime}+1}L^{\prime}_{J^{\prime}}\Big|r\Big|n^{2S+1}L_{J}\right\rangle^2,
\end{eqnarray}
and the magnetic transition width is,
\begin{eqnarray} \label{Eq:M1}
\Gamma^{M_1}&=&\frac{\alpha}{3}\omega_{\gamma}^{3}\delta_{LL^{\prime}}\delta_{SS^{\prime}\pm1}\frac{2J^{\prime}+1}{2L+1}\Big\langle n^{\prime2S^{\prime}+1}L^{\prime}_{J^{\prime}}\Big|\frac{e_{1}}{m_{1}}j_{0}\Big(\omega_{\gamma}\frac{m_{2}r}{m_{1}+m_{2}}\Big)
\nonumber\\
& & {}-\frac{e_{2}}{m_{2}}j_{0}\Big(\omega_{\gamma}\frac{m_{1}r}{m_{1}+m_{2}}\Big)\Big|n^{2S+1}L_{J}\Big\rangle^2, \label{Eq:M1}
\end{eqnarray}
where $\left \langle n^{\prime 2S^{\prime }+1}L^{\prime }_{J^{\prime }}\right|$ and $\left \langle n^{2S+1}L_{J}\right|$ represent the space function of the final and initial states, respectively, which are estimated by the modified GI model. $e_{1}(m_1)$ and $e_{2}(m_2)$ represent the charge (mass) of charm and anti-light quark, respectively. $\alpha$ is the fine structure constant.

\section{Numerical Results and discussions}
 For the heavy-light system, the $^1L_\ell$ and $^3L_\ell$ states have the same $J^P$ quantum numbers, thus the physical states are the mixtures of these two states. In the present work, we have,
\begin{eqnarray}
\left(
\begin{array}{c}
|nL_1 \rangle\\
|nL_1^\prime\rangle	
\end{array}
\right)=
\left(
\begin{array}{cc}
\cos\theta_{L} &\sin \theta_L\\
-\sin\theta_{L} &\cos \theta_L\\
\end{array}
\right)
\left(
\begin{array}{c}
|n^1L_1 \rangle\\
|n^3L_1\rangle	
\end{array}
\right)
\end{eqnarray}
where $L=P$ and $D$ correspond to $P$ and $D$ wave charmed mesons, respectively. In the heavy quark limit, the mixing angle $\theta_{P}$ and $\theta_{D}$ are estimated to be $-54.7^{\circ}$ and $-50.8^{\circ}$, respectively. Besides the above mixing, the tensor term in the Hamiltonian can lead to the mixing between the states with an angular momentum difference of two, such as the mixing between $^3S_1$ and $^3D_1$. As indicated in Refs. \cite{Song:2015fha,Song:2015nia},
such kind of mixing are rather small, thus in the following estimations, such mixing are neglected.

\begin{table}[t]
\caption{The $E_1$ transition width of S-P$\gamma$ process in unit of keV. The results from Ref. \cite{Close:2005se,Godfrey:2015dva} are also presented for comparison.  \label{Tab:S2P}}
\centering
\begin{tabular}{p{1.5cm}<\centering p{1.5cm}<\centering p{1.6cm}<\centering p{1.6cm}<\centering p{1.6cm}<\centering}
\toprule[1pt]
\multirow{2}{*}{Initial State} & \multirow{2}{*}{Final State} & \multicolumn{3}{c}{ Width ($c\bar{u}$/$c\bar{d}$)} \\
\cmidrule[0.5pt]{3-5}
 &  & Present & Ref.~\cite{Close:2005se} & Ref.~\cite{Godfrey:2015dva} \\
 \midrule[0.5pt]
$2^{1}S_{0}$ & $1P_{1}$ & 11.5 / 1.2 & & 44.6 / 4.61 \\
& $1P^{ \prime }_{1}$ & 26.2 / 2.7 & 77.8 / 4.7 & 7.83 / 0.809 \\
$2^{3}S_{1}$ & $1^{3}P_{0}$ & 21.6 / 2.2 & 66 / 3.8 & 23.6 / 2.43 \\
& $1P_{1}$ & 21.5 / 2.2 & 40.7 / 2.3 & 8.25 / 0.852 \\
 & $1P^{ \prime }_{1}$ & 11.8 / 1.2 & & 29.8 / 3.08 \\
 & $1^{3}P_{2}$ & 35.8 / 3.7 & 163 / 9.4 & 45.5 / 4.70 \\
 \midrule[0.5pt]
$3^{1}S_{0}$  & $1P_{1}$ & 11.4 / 1.2 & & 39.3 / 4.06 \\
& $1P^{ \prime }_{1}$ & 23.2 / 2.4 & & 8.64 / 0.89 \\
 & $2P_{1}$ & 22.0 / 2.3 & & 138 / 14.3 \\
 & $2P^{ \prime }_{1}$ & 68.1 / 7.0 & & 18.7 / 1.93 \\
$3^{3}S_{1}$ & $1^{3}P_{0}$ & 2.4 / 0.3 & & 2.60 / 0.269 \\
 & $1P_{1}$ & $\sim0$ / $\sim0$ & & $\sim0$ / $\sim0$ \\
 & $1P^{ \prime }_{1}$ & $\sim0$ / $\sim0$ & & 0.32 / $\sim0$ \\
 & $1^{3}P_{2}$ & 40.7 / 4.2 & & 65.7 / 6.79 \\
 & $2^{3}P_{0}$ & 18.6 / 1.9 & & 24.4 / 2.52 \\
 & $2P_{1}$ & 30.8 / 3.2 & & 24.9 / 2.57 \\
 & $2P^{ \prime }_{1}$ & 21.3 / 2.2 & & 41.7 / 4.30 \\
 & $2^{3}P_{2}$ & 84.9 / 8.8 & & 131 / 13.5 \\
\bottomrule[1pt]
\end{tabular}
\end{table}

\begin{table*}[htbp]
\caption{The same as Table~\ref{Tab:S2P} but for  $P-S\gamma$ processes. The theoretical estimation in Refs. \cite{Godfrey:2015dva,Close:2005se,Colangelo:1993zq,Korner:1992pz} are also presented for comparison. The mixing angles of Ref. \cite{Godfrey:2015dva} and Ref.~\cite{Close:2005se} are $\theta_{1P}=-25.68^{\circ}/\theta_{2P}=-29.39^{\circ}$ and $\theta_{1P}=-26^{\circ}$ respectively, while the mixing angle is ignored in Ref. \cite{Colangelo:1993zq}. \label{Tab:P2S}}
\centering
\begin{tabular}{p{2cm}<\centering p{2cm}<\centering p{2.5cm}<\centering p{2.5cm}<\centering p{2.5cm}<\centering p{2.5cm}<\centering p{2.5cm}<\centering}
\toprule[1pt]
\multirow{2}{*}{Initial State} & \multirow{2}{*}{Final State} & \multicolumn{5}{c}{Width ($c\bar{u}$/$c\bar{d}$)} \\
\cmidrule[0.5pt]{3-7}
 &  & Present & Ref.~\cite{Godfrey:2015dva} & Ref.~\cite{Close:2005se} & Ref.~\cite{Colangelo:1993zq} & Ref.~\cite{Korner:1992pz} \footnote{Only the $E_1$ transitions between the neutral charmed meson were estimated.}\\
 \midrule[0.5pt]
 $1^{3}P_{0}$ & $1^{3}S_{1}$ & 275.4 / 28.5 & 288 / 30 & 304 / 17 & $115\pm54 / <2.8$ & \\
 $1P_{1}$ & $1^{1}S_{0}$ & 254.9 / 26.4 & 640 / 66 & 349.3 / 19.7 & & $245\pm18$\\
 & $1^{3}S_{1}$ & 290.2 / 30.0 & 82.8 / 8.6 & & & $60\pm5$\\
 $1P^{ \prime }_{1}$ & $1^{1}S_{0}$ & 533.1 / 55.1 & 156 / 16.1 & & $14\pm6 / <3.3$ & \\
 & $1^{3}S_{1}$ & 155.4 / 16.1 & 386 / 39.9 & 549.5 / 30.9 & $93\pm44 / <2.3$ & \\
 $1^{3}P_{2}$ & $1^{3}S_{1}$ & 577.4 / 59.7 & 592 / 61.2 & 895 / 51 & & \\
 \midrule[0.5pt]
  $2^{3}P_{0}$ & $1^{3}S_{1}$ & 63.7 / 6.6 & 76.4 / 7.90 & & & \\
   & $2^{3}S_{1}$ & 362.6 / 37.5 & 427 / 44.1 & & &\\
  $2P_{1}$ & $1^{1}S_{0}$ & 7.8 / 0.8 & 14.1 / 1.46 & & & \\
 & $1^{3}S_{1}$ & 7.4 / 0.8 & 3.70 / 0.382 & & &\\
 & $2^{1}S_{0}$ & 180.9 / 18.7 & 384 / 39.6 & & &\\
 & $2^{3}S_{1}$ & 275.5 / 28.5 & 88.9 / 9.19 & & & \\
 $2P^{ \prime }_{1}$ & $1^{1}S_{0}$ & 14.9 / 1.5 & 4.88 / 0.505 & & &\\
          & $1^{3}S_{1}$ & 3.5 / 0.4 & 12.9 / 1.34 & & &\\
          & $2^{1}S_{0}$ & 315.2 / 32.6 & 162 / 16.7 & & &\\
          & $2^{3}S_{1}$ & 117.0 / 12.1 & 396 / 40.9 & & &\\
  $2^{3}P_{2}$ & $1^{3}S_{1}$ & 14.0 / 1.4 & 10.9 / 1.12 & & &\\
   & $2^{3}S_{1}$ & 366.9 / 37.9 & 425 / 43.9 & & &\\
\bottomrule[1pt]
\end{tabular}
\end{table*}

\begin{table}[htbp]
\caption{The same as Table~\ref{Tab:S2P} but for  of $P\to D\gamma$ processes.\label{Tab:P2D}. The mixing angles of Ref. \cite{Godfrey:2015dva} is $\theta_{1D}=-38.17^{\circ}$}
\centering
\begin{tabular}{p{2cm}<\centering p{2cm}<\centering p{2cm}<\centering p{2cm}<\centering}
\toprule[1pt]
Initial State & Final State & \multicolumn{2}{c}{Width ($c\bar{u}$/$c\bar{d}$)} \\
\cmidrule[0.5pt]{3-4}
 &  & Present & Ref.~\cite{Godfrey:2015dva} \\
 \midrule[0.5pt]
 $2^{3}P_{0}$ & $1^{3}D_{1}$ & 19.8 / 2.0 & 30.5 / 3.15 \\
 $2P_{1}$  & $1^{3}D_{1}$ & 2.5 / 0.3 & 1.26 / 0.13 \\
 & $1D_{2}$ & 0.9 / 0.1 & 25.1 / 2.59 \\
 & $1D^{ \prime }_{2}$ & 18.8 / 1.9 & 0.476 / $\sim0$\\
 $2P^{ \prime }_{1}$  & $1^{3}D_{1}$ & 3.1 / 0.3 & 9.30 / 0.961 \\
 & $1D_{2}$ & 14.0 / 1.4 & 0.385 / $\sim0$\\
 & $1D^{ \prime }_{2}$ & 1.4 / 0.1 & 26.7 / 2.76 \\
 $2^{3}P_{2}$  & $1^{3}D_{1}$ & 0.2 / $\sim 0$ & 0.36 / $\sim0$ \\
 & $1D_{2}$ & 2.2 / 0.2 & 2.02 / 0.208 \\
 & $1D^{ \prime }_{2}$ & 1.2 / 0.1 & 2.47 / 0.255 \\
 & $1^{3}D_{3}$ & 23.8 / 2.5 & 34.2 / 3.53 \\
\bottomrule[1pt]
\end{tabular}
\end{table}

\begin{table}[htbp]
\caption{The same as Table \ref{Tab:S2P} but for $D\to P\gamma$ processes.\label{Tab:D2P}}
\centering
\begin{tabular}{p{2cm}<\centering p{2cm}<\centering p{2cm}<\centering p{2cm}<\centering}
\toprule[1pt]
Initial State & Final State & \multicolumn{2}{c}{Width ($c\bar{u}$/$c\bar{d}$)} \\
\cmidrule[0.5pt]{3-4}
 &  & Present & Ref.~\cite{Godfrey:2015dva} \\
 \midrule[0.5pt]
 $1^{3}D_{1}$ & $1^{3}P_{0}$ & 488.9 / 50.5 & 521 / 53.8 \\
  & $1P_{1}$ & 177.0 / 18.3 & 55.9 / 5.78 \\
 & $1P^{ \prime }_{1}$ & 174.2 / 18.0 & 222 / 22.9 \\
 & $1^{3}P_{2}$ & 14.4 / 1.5 & 15.9 / 1.64 \\
 $1D_{2}$  & $1P_{1}$ & 21.0 / 2.2 & 642 / 66.4 \\
 & $1P^{ \prime }_{1}$ & 579.4 / 59.9 & 12.2 / 1.26 \\
 & $1^{3}P_{2}$ & 88.3 / 9.1 & 55.4 / 5.72 \\
 $1D^{ \prime }_{2}$ & $1P_{1}$ & 571.4 / 59.0 & 64.7 / 6.69 \\
 & $1P^{ \prime }_{1}$ & 77.6 / 8.0 & 640 / 66.1 \\
 & $1^{3}P_{2}$ & 60.9 / 6.3 & 116 / 11.9 \\
  $1^{3}D_{3}$ & $1^{3}P_{2}$ & 625.0 / 64.6 & 686 / 70.9 \\
 \midrule[0.5pt]
  $2^{3}D_{1}$ & $1^{3}P_{0}$ & 8.2 / 0.9 \\
 & $1P^{ \prime }_{1}$ & 0.7 / 0.1 \\
 & $1P_{1}$ & 1.5 / 0.1 \\
 & $1^{3}P_{2}$ & 3.6 / 0.4 \\
 & $2^{3}P_{0}$ & 314.9 / 32.5 \\
 & $2P^{ \prime }_{1}$ & 76.6 / 7.9 \\
 & $2P_{1}$ & 172.2 / 17.8 \\
 & $2^{3}P_{2}$ & 15.5 / 1.6 \\
$2D_{2}$ & $1P_{1}$ & 0.7 / 0.1 \\
& $1P^{ \prime }_{1}$ & 5.9 / 1.6 \\
& $1^{3}P_{2}$ & 4.9 / 0.5 \\
& $2P_{1}$ & 15.2 / 1.6 \\
& $2P^{ \prime }_{1}$ & 435.8 / 45.0 \\
& $2^{3}P_{2}$ & 75.1 / 7.7 \\
 $2D^{ \prime }_{2}$ & $1P_{1}$ & 6.1 / 0.6 \\
 & $1P^{ \prime }_{1}$ & 0.3 / $\sim0$ \\
 & $1^{3}P_{2}$ & 5.1 / 0.5 \\
 & $2^{3}P_{0}$ & 77.6 / 8.0 \\
  & $2P_{1}$ & 391.8 / 40.5 \\
 & $2P^{ \prime }_{1}$ & 57.3 / 5.9 \\
 & $2^{3}P_{2}$ & 52.9 / 5.4 \\
 $2^{3}D_{3}$ & $1^{3}P_{2}$ & 1.1 / 0.1 \\
& $2^{3}P_{2}$ & 442.4 / 45.7 \\
 \bottomrule[1pt]
\end{tabular}
\end{table}

\subsection{Electric transitions}
As indicated in Eq. (\ref{Eq:E1}), the $E_1$ transitions occur between the states with the same spins but angular momenta difference of one. In Tables \ref{Tab:S2P}-\ref{Tab:D2P}, we present the electric transitions for $S\to P \gamma$, $P\to S\gamma$, $P\to D \gamma$, and $D\to P\gamma $, respectively, where $S$, $P$, and $D$ refer to the $S-$, $P-$ and $D-$wave charmed mesons, respectively. It is worthwhile to mention that in the quark model, the masses of the up and down quark are usually in the same value, and then the estimated mass spectra and wave functions for the charged and neutral charmed mesons are identical. Thus for the $E_1$ transition as shown in Eq. (\ref{Eq:E1}), the only difference for the charged and neutral charmed mesons is the charge of the involved quarks. Taking $m_c=1628$ MeV and $m_u=m_d=220$ MeV~\cite{Godfrey:1985xj}, one can find the E1 transition for the neutral states is about 9.7 times of the corresponding one for the charged states.

\begin{itemize}[leftmargin=*]
\item{\it $S \to P\gamma$ processes.} In Table \ref{Tab:S2P}, we present our estimations of the electric transition widths from $S$ wave to $P$ wave charmed mesons. where $c\bar{u}$ and $c\bar{d}$ refer to the neutral and charged charmed mesons, respectively. For comparison, we also present the results from Ref.~\cite{Close:2005se, Godfrey:2015dva}. In Ref.~\cite{Close:2005se} the wave functions were evaluated in a simple nonrelativistic quark model with color Coulomb plus linear scalar confinement interaction with addition of a Gaussian smeared contact hyperfine interaction term. For the $2S \to 1P \gamma $ processes, the results of Ref.~\cite{Close:2005se} are several times larger than those in the present work due to the different model parameters. In Ref.~\cite{Godfrey:2015dva}, the radiative decays were investigated in the GI model. From the table, one can find that the radiative transition widths for the process involving $P_{1}/P_1^{\prime}$ states are much different due to the different mixing angles used in the present estimations and in Ref.~\cite{Godfrey:2015dva}, while the widths for other channels are very similar. The influence of mixing angle on the radiative decay process will be discussed later. Moreover, the widths of $D(2^{1}S_{0})^{0}$ and $D^{*}(2^{3}S_{1})^{0}$ were measured to be  $199\pm5\pm17$ MeV and $149\pm4\pm20$ MeV respectively \cite{LHCb:2019juy}, thus the branching ratios of $2S \to 1P \gamma$ processes are of order $10^{-4}$ and $10^{-5}$ for neutral and charged charmed mesons, respectively.

As for $3S$ states, our estimation indicate that the partial widths of $3S \to 1P \gamma $ processes are suppressed comparing to those of $3S\to 2P \gamma$ processes due to the node effects. In particular, our estimations indicate that the widths for $D^0(3^1S_0) \to D_1^{\prime 0}(2P^\prime_1) \gamma$ and $D(3^3S_1)^0 \to D_2(2^3P_2)^{0} \gamma$ are nearly 100 keV. In Ref.~\cite{Song:2015fha}, the widths of $D(3^1S_0)$ and $D^{\ast}(3^3S_1)$ were about 100 MeV, thus the branching ratios of  $D^0(3^1S_0) \to D_1^{\prime 0}(2P^\prime_1) \gamma$ and $D^{\ast 0}(3^3S_1) \to D_2^0(2^3P_2) \gamma$ should be of order $10^{-3}$, which may be detected by further measurements in LHCb and Belle II.

\item{\it $P \to S\gamma$ processes.} In Table~\ref{Tab:P2S}, we present our estimations of $E_1$ transitions width of  $P-S\gamma$ processes. The estimations in Refs. \cite{Godfrey:2015dva, Close:2005se, Colangelo:1993zq, Korner:1992pz} are also presented for comparison. For $1^3P_0 \to 1^3S_{1} \gamma$ and $1^3P_2 \to 1^3S_1 \gamma$ processes, we find our results are almost the same as those in Ref.~\cite{Godfrey:2015dva} since the modified relativistic quark model used in the present work are similar to the GI model employed in Ref.~\cite{Godfrey:2015dva}. For $1P_1/1P_1^\prime \to 1S \gamma$, our estimation is a bit different with those in  Ref.~\cite{Godfrey:2015dva} due to the different mixing angle $\theta_{1P}$ used in the estimations. 

To explore the influence of the mixing angle on the electrical transition width, we present the mixing angle dependences of the radiative decay widths of the process $D_1^0(1P_1) \to D^{(\ast) 0} \gamma$ and $D_1^{\prime 0}(1P_1^\prime) \to D^{(\ast) 0} \gamma$  in Figs.~\ref{Fig:His2} and \ref{Fig:His3}, respectively. As for the electric transition processes, the initial and final states should have the same spin, thus, the widths of $D^{0}_{1}(1P_{1}) \to D^{*0}\gamma$ and $D^{\prime 0}_{1}(1P^{\prime}_{1}) \to D^{0}\gamma$ should be proportional to $\sin^2 \theta_{1P}$, while the widths of $D^{\prime 0}_{1}(1P^\prime_{1}) \to D^{*0}\gamma$ and $D^{ 0}_{1}(1P_{1}) \to D^{0}\gamma$ are proportional to $\cos^2 \theta_{1P}$. In the figures, we also shows the different mixing angle used in the present work and in Ref.~\cite{Godfrey:2015dva}, which are $\theta_{1P}=-54.7^{\circ}$ and $\theta_{1P}=-25.68^{\circ}$, respectively. As indicated in the figures, one can find the widths for the processes of $D^{0 (\prime)}_{1}(1P^{(\prime)}_{1}) \to D^{(\ast)0}\gamma$ calculated in the present work  result is similar to those in Ref.~~\cite{Godfrey:2015dva} when one take the same mixing angle $\theta_{1P}$

    Our estimations are also comparable to those in Ref.~\cite{Close:2005se}, where the wave functions are estimated in a non-relativistic quark model. However, the estimations based heavy quark effective theory in Refs.~\cite{Colangelo:1993zq, Korner:1992pz} are several times smaller than the potential model estimation in the present work and in Refs.~\cite{Godfrey:2015dva, Close:2005se}. For $1P \to 1S \gamma$ processes, the present estimations indicate the widths of all the possible processes are above 100 keV. The PDG average of the widths of $D_0(1^3P_0)$, $D_1(1P_1)$, $D_1(1P_1^\prime)$, and $D_2(1^3P_2)$ are $229 \pm 16$, $31.3 \pm 1.9$, $314\pm 29$, and $47.3 \pm 0.8$ MeV, respectively. Thus, for $D_1(1P_1)$ and $D_2(1^3P_2)$, the branching ratios of the $E_1$ transitions are of order of $10^{-3} \sim 10^{-2}$, which are large enough to be detected.

\begin{figure}[htb]
\centering
\includegraphics[width=80mm]{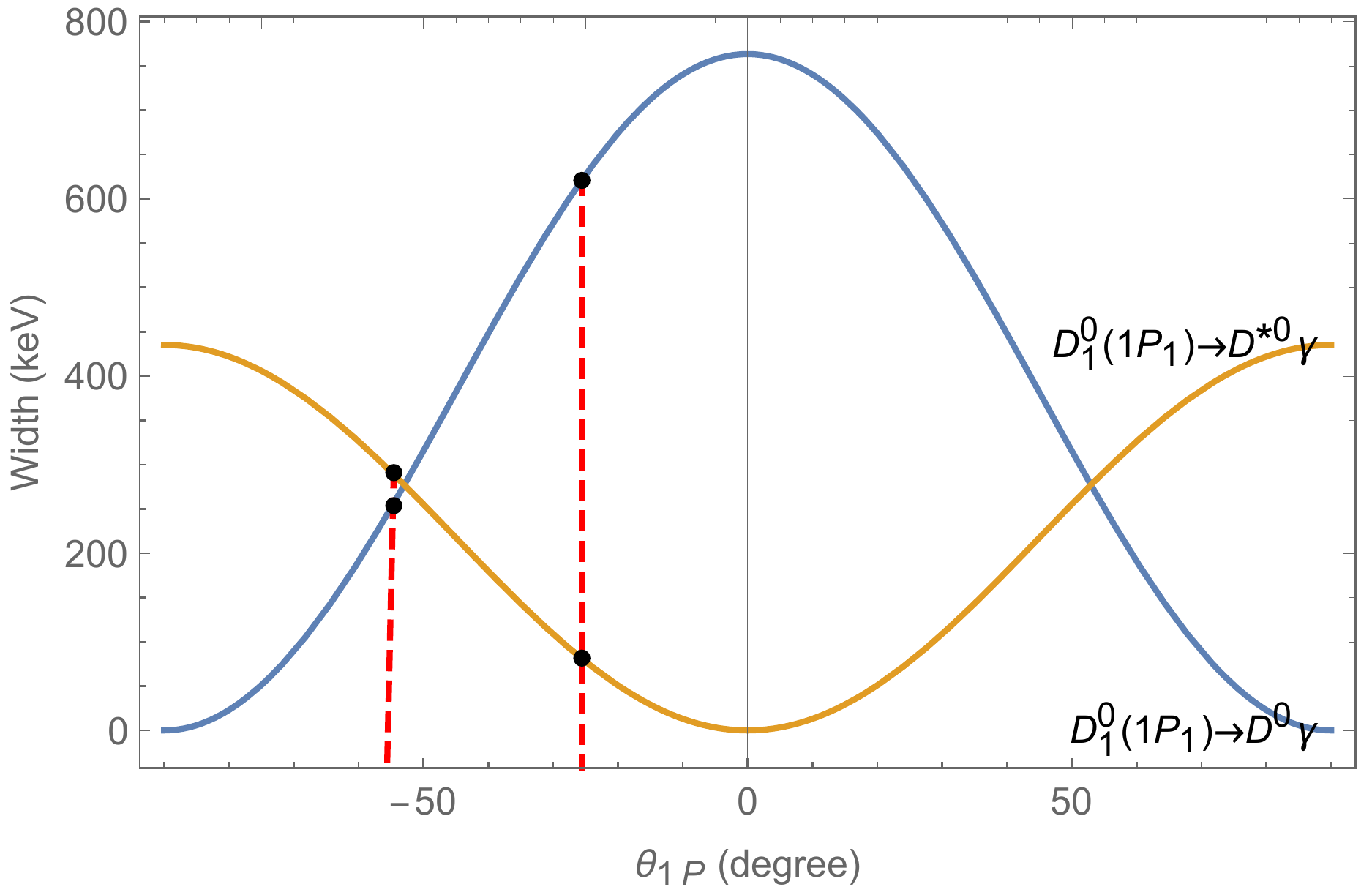}
\caption{The width of the process $D^{0}_{1}(1P_{1}) \to D^{*0}/D^{0}\gamma$ depending on the mixing angle $\theta_{1P}$. \label{Fig:His2}}
\end{figure}

\begin{figure}[htb]
\centering
\includegraphics[width=80mm]{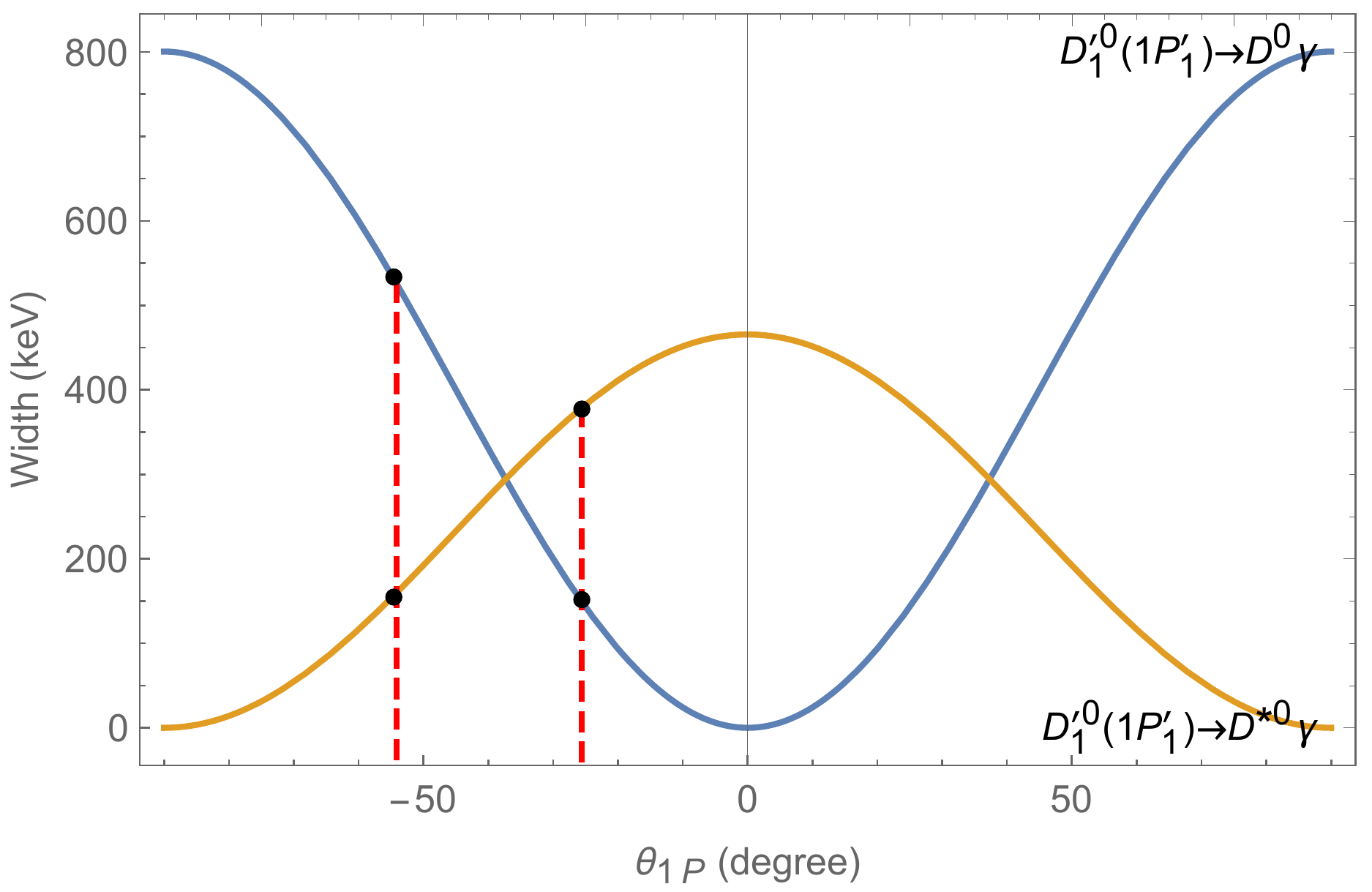}
\caption{The same as Fig.~\ref{Fig:His2} but for $D^{\prime 0}_{1}(1P^{\prime}_{1}) \to D^{*0}/D^{0}\gamma$. \label{Fig:His3}}
\end{figure}

Similar to the case of $S \to P \gamma$ process, our estimation also indicate $2P \to 1S \gamma$ processes are suppressed comparing to $2P \to 2S \gamma$ process due to node effects. In particular, we find widths of $D_0^0(2^3P_0) \to D^{\ast 0}(2^3S_1) \gamma$,  $D_1^0(2P_1) \to D^0(2^1S_0) \gamma$,  $D_1^0(2P_1) \to D^{\ast 0}(2^3S_1) \gamma$,  $D_1^{\prime 0}(2P_1^\prime) \to D^0(2^1S_0) \gamma$,  $D_1^{\prime 0}(2P_1^\prime) \to D^{\ast 0}(2^3S_1) \gamma$, and $D_2^0(2^3P_2) \to D^{\ast 0}(2^3S_1) \gamma$ are larger than 100 keV. Our estimations in Ref.~\cite{Song:2015fha} indicated that the widths of $2P$ charmed mesons to be about 100 MeV, thus, the branching ratios of the above processes should be of order $10^{-3}$.

\item{\it $P \to D\gamma$ processes.} As shown in Table \ref{Tab:P2D}, our estimations indicate the $E_1$ transition widths for $2P \to 1D \gamma$ are rather small, which are similar to those in  Ref.~\cite{Godfrey:2015dva}. In this work only the widths of $D_0^0(2^3P_0) \to D_1^0(1^3D_1) \gamma$, $D_1^0(2P_1) \to D_2^{\prime 0}(1D^\prime_2) \gamma$, $D_1^{\prime 0}(2P_1^\prime) \to D_2^{0}(1D_2) \gamma$, and $D_2^0(2^3P_2) \to D_3^0(1^3D_3) \gamma$ are several tens keV. However, the widths of $2P$ charmed mesons are about 100 MeV, thus, the branching ratio of these channel should be at most of the order of $10^{-4}$. As for other seven channels, their width are ever smaller. Thus, it is hard to detect the $2P \to 1D \gamma$ processes experimentally.

\item{\it $D\to P\gamma$ processes.}
Our estimations of the widths of $D\to P \gamma$ are listed in Table \ref{Tab:D2P}. By comparing our estimations with the results in Ref.~\cite{Godfrey:2015dva}, one can find the transition widths are very similar, except the processes with mixing states, such as $1P_1$ and $1P_1^\prime$. As indicated in Fig.~\ref{Fig:His2} and Fig.~\ref{Fig:His3}, such discrepancies are resulted from the different mixing angles adopted in the calculations. For $1D \to P \gamma $ processes, our estimations indicate that the widths of $D_1^0(1^3D_1) \to D_0^0(1^3P_0) \gamma$, $D_2^0(1D_2) \to D_1^{\prime 0}(1P^\prime_1) \gamma$, $D_2^{\prime 0}(1D_2^\prime) \to D_1^{0}(1P_1) \gamma$, and $D_3^0(1D_3) \to D_2^0(1^3P_2) \gamma$ are more than 400 keV. The experimental measurement and theoretical estimations in Ref.~\cite{Song:2015fha} indicated that the widths of $D_1(1^3D_1)$, $D_2(1D_2)$ and $D_2^\prime (1D_2^\prime)$ are about 100 MeV, thus, the branching ratios of $D_1^0(1^3D_1) \to D_0^0(1^3P_0) \gamma$, $D_2^0(1D_2) \to D_1^{\prime 0}(1P_1^\prime) \gamma$, and $D_2^{\prime 0}(1D_2^\prime) \to D_1^{0}(1P_1) \gamma$ should be of the order of $10^{-3}$. As for $D_3(1D_3)$, the width is estimated to be 18.3 MeV, which indicate the branching ratio of $D_3^0(1D_3) \to D_2^0(1^3P_2) \gamma$ is about $3\%$, which should be observable experimentally.

As for electric decays of $2D$ states, we find the widths of $2D \to 1P \gamma$ are smaller than the corresponding one of $2D \to 2P \gamma$, which is similar to the case of the electric decays of $3S$ and $2P$ states. For the $2D\to 2P \gamma$ process, our estimations show that the widths of the $D_1^0(2^3D_1) \to D_0^0(2^3P_0) \gamma$, $D_2^0(2D_2) \to D_1^{\prime 0}(2P_1^\prime) \gamma$, $D_2^{\prime 0}(2D_2^\prime)\to D_1^{0}(2P_1) \gamma$, and $D_3^0(2^3D_3) \to D_2^0(2^3P_2) \gamma$ processes are 314.8, 435.8, 391.8, and 442.4 keV, respectively. As indicated in Ref.~\cite{Song:2015fha}, the widths of $D(2^3D_1)$ and $D(2D_2)$ are nearly 100 MeV, while the widths of $D_2^\prime (2D_2^\prime)$ and $D_3(2^3D_3)$ are about 30 MeV, then, one can find the branching ratios of $D_1^0 (2^3D_1)\to D_0^0(2^3P_0) \gamma$ and $D_2^0(2D_2) \to D_1^{\prime 0}(2P_1^\prime) \gamma$ are of the order of $10^{-3}$, while the branching ratios of $D_2^{\prime 0}(2D_2^\prime) \to D_1^{0}(2P_1) \gamma$ and $D_3^0(2^3D_3) \to D_2^0(2^3P_2) \gamma$ are of the order of $10^{-2}$, which should be accessible in further experiments.
	
\end{itemize}

\begin{table*}[htbp]
\caption{The $M_1$ transition widths of $S-S\gamma$ processes in unit of keV. For comparison, we also present the results from Refs.~\cite{Godfrey:2015dva,Close:2005se,Colangelo:1993zq}. \label{Tab:S2S}}
\centering
\begin{tabular}{p{2.4cm}<\centering p{2.4cm}<\centering p{2.9cm}<\centering p{2.9cm}<\centering p{2.9cm}<\centering p{2.9cm}<\centering}
\toprule[1pt]
\multirow{2}{*}{Initial State} & \multirow{2}{*}{Final State}& \multicolumn{4}{c}{Width ($c\bar{u}$/$c\bar{d}$ )} \\
\cmidrule[0.5pt]{3-6}
 &  & Present  & Ref.~\cite{Godfrey:2015dva}  & Ref.~\cite{Close:2005se}   & Ref.~\cite{Colangelo:1993zq}  \\
 \midrule[0.5pt]
 $1^{3}S_{1}$ & $1^{1}S_{0}$ & 97.0 / 9.9 & 106/10.2 & 32 / 1.8 & 43.6 / 1.1\\
 \midrule[0.5pt]
 $2^{1}S_{0}$ & $1^{3}S_{1}$ & 0.4 / 4.3 & $\sim0$ / 5.80 & & \\
 $2^{3}S_{1}$ & $1^{1}S_{0}$ & 526.4 / 87.0 &600 / 100 & & \\
  & $2^{1}S_{0}$ & 5.4 / 0.5 & 6.26 / 0.641 & & \\
  \midrule[0.5pt]
  $3^{1}S_{0}$ & $1^{3}S_{1}$ & 0.8 / 3.6 & $\sim0$ / 6.27 & & \\
  & $2^{3}S_{1}$ & 24.6 / 13.2 & 44.5 / 22.4  & & \\
  $3^{3}S_{1}$ & $1^{1}S_{0}$ & 547.9 / 97.2 & 663 / 119 & & \\
  & $2^{1}S_{0}$ & 189.7 / 68.0 & 257 / 48.2 & & \\
  & $3^{1}S_{0}$ & 1.5 / 0.1 & 2.03 / 0.208 & & \\
 \bottomrule[1pt]
\end{tabular}
\end{table*}

\begin{table}[htbp]
\caption{The same as Table \ref{Tab:S2S} but for  $P\to P\gamma$ processes.\label{Tab:P2P}}
\centering
\begin{tabular}{p{2cm}<\centering p{2cm}<\centering p{4cm}<\centering}
\toprule[1pt]
InitialState & FinalState & \multicolumn{1}{c}{Width ($c\bar{u}$/$c\bar{d}$ )} \\
 \midrule[0.5pt]
 $1P_{1}$ & $1^{3}P_{0}$ & 1.4 / 0.1 \\
 $1P^{ \prime }_{1}$ & $1P_{1}$ & $\sim0$ / $\sim0$ \\
 $1^{3}P_{2}$ & $1P_{1}$ & 1.4 / 0.1 \\
 & $1P^{ \prime }_{1}$ & 0.5 / 0.1 \\
 \midrule[0.5pt]
 $2^{3}P_{0}$ & $1P_{1}$ & 4.2 / 2.6 \\
 & $1P^{ \prime }_{1}$ & 1.8 / 1.2 \\
 $2P_{1}$   & $1^{3}P_{0}$ & 30.8 / 5.4 \\
 & $1P_{1}$  & 33.5 / 8.2 \\
 & $1P^{ \prime }_{1}$ & 6.0 / 1.2 \\
 & $1^{3}P_{2}$ & 3.7 / 2.9 \\
 & $2^{3}P_{0}$ & $\sim0$ / $\sim0$ \\
 $2P^{ \prime }_{1}$ & $1^{3}P_{0}$ & 17.6 / 3.1 \\
 & $1P_{1}$ & 7.3 / 1.4 \\
 & $1P^{ \prime }_{1}$ & 34.9 / 8.5 \\
 & $1^{3}P_{2}$ & 0.6 / 0.4 \\
 & $2^{3}P_{0}$ & $\sim0$ / $\sim0$ \\
 & $2P_{1}$ & $\sim0$ / $\sim0$ \\
 $2^{3}P_{2}$ & $1P_{1}$ & 37.6 / 6.5 \\
 & $1P^{ \prime }_{1}$ & 78.5 / 13.6 \\
 & $2P_{1}$ & 0.2 / $\sim0$ \\
 & $2P^{ \prime }_{1}$ & $\sim0$ / $\sim0$ \\
   \bottomrule[1pt]
\end{tabular}
\end{table}

\begin{table}[htbp]
\caption{The same as Table \ref{Tab:S2S} but for  $D\to D\gamma$ processes.\label{Tab:D2D}}
\centering
\begin{tabular}{p{2cm}<\centering p{2cm}<\centering p{4cm}<\centering}
\toprule[1pt]
InitialState & FinalState & \multicolumn{1}{c}{Width ($c\bar{u}$/$c\bar{d}$ )} \\
 \midrule[0.5pt]
$1D_{2}$ & $1^{3}D_{1}$ & $\sim0$ / $\sim0$ \\
$1D^{ \prime }_{2}$ & $1^{3}D_{1}$ & 0.1 / $\sim0$ \\
& $1D_{2}$ & $\sim0$ / $\sim0$ \\
$1^{3}D_{3}$ & $1D_{2}$ & $\sim0$ / $\sim0$ \\
\midrule[0.5pt]
$2^{3}D_{1}$ & $1D_{2}$ & 2.3 /1.2 \\
& $1D^{ \prime }_{2}$ & 3.0 / 1.5 \\
& $2D_{2}$ & $\sim0$ / $\sim0$ \\
$2D_{2}$ & $1^{3}D_{1}$ & 12.1 / 2.1 \\
& $1D_{2}$ & 21.1 / 5.0 \\
& $1D^{ \prime }_{2}$ & 0.3 / 0.1 \\
& $1^{3}D_{3}$ & 4.8 / 1.9 \\
$2D^{ \prime }_{2}$ & $1^{3}D_{1}$ & 20.0 / 3.4 \\
& $1D_{2}$ & 1.9 / 0.3 \\
& $1D^{ \prime }_{2}$ & 21.5 / 5.0 \\
& $2^{3}D_{1}$ & $\sim0$ / $\sim0$ \\
& $2D_{2}$ & $\sim0$ / $\sim0$ \\
$2^{3}D_{3}$ & $1D_{2}$ & 17.5 / 3.1 \\
& $2D_{2}$ & $\sim0$ / $\sim0$ \\
 \bottomrule[1pt]
\end{tabular}
\end{table}

\subsection{Magnetic transitions}

As indicated in Eq.~(\ref{Eq:M1}), The $M_1$ transitions occur  between the states with the same angular but different spin. Our estimation of the $M_1$ transition for $S\to S \gamma$, $P\to P \gamma$ and $D\to D\gamma$ are listed in Table \ref{Tab:S2S}-\ref{Tab:D2D}, respectively.

\begin{itemize}[leftmargin=*]

\item{ \it $S\to S \gamma$ processes.} In Table~\ref{Tab:S2S}, we present our estimations to the $M_1 $ transitions between two $S$ wave charmed mesons. For comparison, we also present the estimations from GI model~\cite{Godfrey:2015dva}, nonrelativistic quark model~\cite{Close:2005se}, and heavy quark effective theory~\cite{Colangelo:1993zq}. As for $D^\ast \to D \gamma$, we find the width for $D^{\ast 0} \to D^0 \gamma $ and $D^{\ast +} \to D^+ \gamma $ are about 97 and 9.9 keV, respectively, which are consistent with the estimation in GI model and several times larger than the one in non-relativistic quark model~\cite{Close:2005se} and heavy quark effective theory~\cite{Colangelo:1993zq}.

On the experimental side, the widths of $D^{\ast+}$ is $83.4 \pm 1.8$ keV ,and the branching ratio of $D^{\ast +} \to D^{+} \gamma$ is measured to be $(1.6 \pm 0.4)\%$, then the measured width of $D^{\ast +} \to D^{+} \gamma$ is $1.33 \pm 0.33 $ keV, which is several times smaller than the estimation in the present work. As for $D^{\ast 0} \to D^0 \gamma $, the  branching ratio has been well determined, which is $(35.3\pm 0.9)\%$, but only the upper limit of the width of $D^{\ast 0}$ is measured, which is 2.1 MeV. Thus, the upper limit of width of $D^{\ast 0} \to D^0 \gamma$ is about 740 keV, which is far above the present estimation. Furthermore, one can roughly estimate the partial widths of $D^{\ast 0} \to D^0 \gamma$ by the relative fraction of $D^{\ast 0} \to D^0 \gamma$ and $D^{\ast 0} \to D^0 \pi^0$. The PDG average of the branching ratio of $D^{\ast 0} \to D^0 \pi^0$ is $(64.7 \pm 0.9 )\%$, which is about $1.83$ times of the one of $D^{\ast 0} \to D^0 \gamma^0$, while  the partial widths of $D^{\ast 0} \to D^0 \pi^0$ can be deduced from the one of $D^{\ast +} \to D^{+} \pi^0$ by considering the isospin symmetry \cite{Dong:2008gb, Chen:2010re}. Then the width of $D^{\ast 0} \to D^0\gamma $ is estimated to be around 15 keV, which is also several times smaller than the present estimation and the results in the GI model.

It should be notice that the magnetic transition width in Eq. (\ref{Eq:M1}) is sensitive to the quark mass. Taking the process $D^\ast \to D \gamma$ as an example, one can expand the spherical Bessel function $j_0(x) \simeq 1 -x^2/6 +\mathcal{O}(x^4)$, then in the leading order approximation, the magnetic transition width is proportional to $(e_1m_2-e_2m_1)^2/(m_1^2m_2^2)$. In the  present modified GI model and GI model, the mass of the light quark is 220 MeV, while in the non-relativistic quark model, the mass of the light quark is usually taken to be 330 MeV. A smaller light quark mass will usually lead to a larger magnetic transition widths. In Fig.~\ref{Fig:His4}, we present the widths of the process $D^{+*} \to D^{+}\gamma$ depending on the light quark mass, where one can find the width will be about 2 keV when we take $m_d=0.33$ GeV, which is several times smaller than the one when taking $m_d=0.22$ GeV and consistent with the experimental measurements.

\begin{figure}[htb]
\centering
\includegraphics[width=80mm]{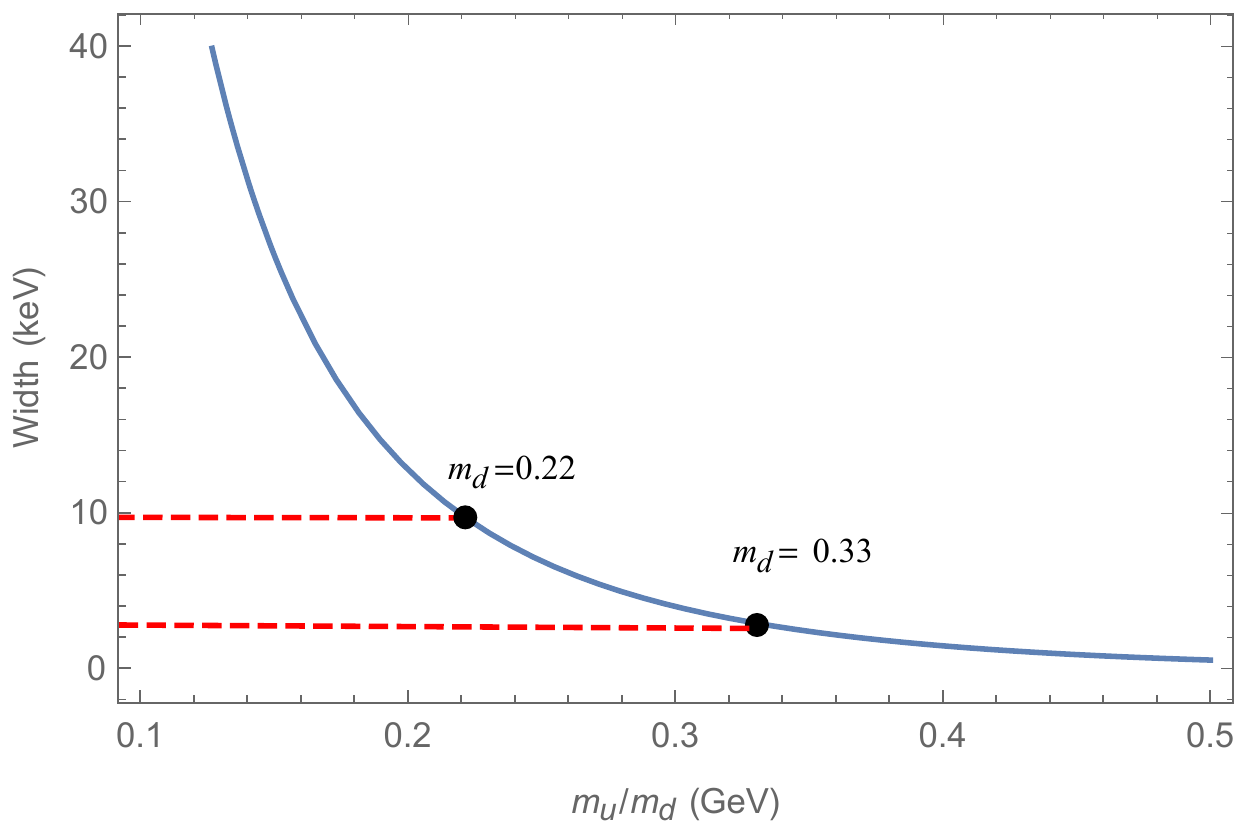}
\caption{The width of $D^{+*} \to D^{+}\gamma$ depending on the light quark moass. \label{Fig:His4}}
\end{figure}

Our estimations indicate that the widths of $D^{\ast 0} (2^3S_1) \to D^{0} \gamma$ and  $D^{\ast 0} (3^3S_1) \to D^{0} \gamma$ are greater than 500 keV, which are also consistent with the estimations in the GI model~\cite{Godfrey:2015dva}. The PDG average of the widths of $D_1^\ast(2600)$ is $141 \pm 23$ MeV, thus the branching ratio of $D^{\ast 0} (2^3S_1) \to D^{0} \gamma$ is about $3.7 \pm 10^{-3}$. As for $D^{\ast 0} (3^3S_1)$, its width was estimated to be 80 MeV in the modified GI model~\cite{Song:2015fha}, then the branching ratio of $D^{\ast 0} (3^3S_1) \to D^{0} \gamma$ is about $6.8 \times 10^{-3}$.

\item{\it $P \to P \gamma$ processes.} As shown in Table \ref{Tab:P2P}, the widths of most of the $P\to P \gamma$ processes are much small. In particular, the widths for $1P \to 1P \gamma$ process, the width are several keV or less than 1 keV. As for $2P \to 1P \gamma$ process, the widths varies from several keV to several tens keV. For $2P \to 2P \gamma $ processes, the widths are very small and most of them are close to 0.

\item{\it $D \to D \gamma$ processes.}  In Table \ref{Tab:D2D}, we present our estimations of the widths of $D\to D \gamma$ process. As for $1D \to 1D \gamma$ and $2D\to 2D \gamma$ processes, the widths are very small and most of them are less than 0.1 keV. As for $2D\to 1D \gamma$ process, the widths are estimated from several keV to several tens keV. The widths of $2D\to 1D \gamma$ are slightly higher than $1D \to 1D \gamma$ and $2D\to 2D \gamma$ processes, but the branching ratio of these channel should be at most of the order of $10^{-4}$.

\end{itemize}

\section{Summary}
In the past two decades, a growing number of charmed mesons have been observed experimentally, which makes the charmed meson spectrum abundant. However, how to categorize these newly observed charmed mesons are also a great challenge for theorists. As one of the most successful  QCD inspired quark models, the GI model can describe the low lying mesons quite well but fail for higher excited states due to the coupled channel effects. In Ref.~\cite{Song:2015fha,Song:2015nia}, we introduced a  screened potential in the GI model to substitute for the coupled channels effects, and in this modified GI model, the higher excited charmed mesons, including the mass spectra and strong decays, can be better described than the GI model. Similar to the strong decays, the radiative transitions can also probe the internal charge structure of mesons and be useful in determining meson structure, thus in the present work, we extend our estimations in Ref.~\cite{Song:2015fha,Song:2015nia} to investigate the radiative decays of the charmed mesons.

By using the wave function obtained from modified GI model, we estimate the radiative transitions between the $mS$, $nP$ and $nD$ charmed mesons ($m=\{1,2,3\}, \ \ n=\{1,2\}$) in the present work.
Our estimations indicate the branching ratios of some processes, including $D_2^0(1^3P_2) \to D^{\ast 0 }(1^3S_1) \gamma$,  $D_3^0(1D_3) \to D_2^0(1^3P_2) \gamma$, $D_2^{\prime 0}(2D_2^\prime) \to D_1^{0}(2P_1) \gamma$, $D_3^0(2^3D_3) \to D_2^0(2^3P_2) \gamma$, and $D^{\ast 0}(1^3S_1) \to D^0(1^1S_0) \gamma$, are of order of $10^{-2}$, which should be sizable to be detected experimentally. Moreover, the branching ratios of some channels, for example, $D_1^0(1P_1) \to D^0(1^1S_0) \gamma$, $D^0(3^1S_0) \to D_1^{\prime 0}(2P_{1}^\prime) \gamma$ and $D^0 (3^3S_1) \to D_2^0(2^3P_2) \gamma$ are estimated to be about $10^{-3}$, which may also be accessible with the accumulation of data in future experiments.

\section*{Acknowledgement}
This work is supported by the National Natural Science Foundation of China (NSFC) under Grant Nos.11775050 and 12175037.

\end{document}